%% file: main.tex
\documentclass[%
 reprint,
 superscriptaddress,
 twocolumn, 
 amsmath,
 amssymb,
 aps,
]{revtex4-1}

\usepackage[colorlinks = true,
            linkcolor = blue,
            urlcolor  = red,
            citecolor = blue,
            anchorcolor = blue]{hyperref}

\usepackage{glossaries}
\input{acronyms.tex}

\usepackage{graphicx}
\usepackage{dcolumn}
\usepackage{bm}
\usepackage[mathlines]{lineno}
\usepackage{orcidlink}

\usepackage[
  color=orange!40,
  textsize=singlespacetiny,
]{todonotes}

\begin{document}

\preprint{}

\title{Fluctuating charge-density-wave correlations in the three-band Hubbard model}

\author{Peizhi Mai\orcidlink{0000-0001-7021-4547}}
\affiliation{Department of Physics and Anthony J Leggett Institute for Condensed Matter Theory, University of Illinois at Urbana-Champaign, Urbana, Illinois 61801, USA\looseness=-1}

\author{Benjamin Cohen-Stead\orcidlink{0000-0002-7915-6280}}
\affiliation{Department of Physics and Astronomy, The University of Tennessee, Knoxville, Tennessee 37996, USA}
\affiliation{Institute of Advanced Materials and Manufacturing, The University of Tennessee, Knoxville, Tennessee 37996, USA\looseness=-1} 

\author{Thomas A. Maier\orcidlink{0000-0002-1424-9996}}
\affiliation{Computational Sciences and Engineering Division, Oak Ridge National Laboratory, Oak Ridge, Tennessee, 37831-6494, USA\looseness=-1}

\author{Steven Johnston\orcidlink{0000-0002-2343-0113}}
\affiliation{Department of Physics and Astronomy, The University of Tennessee, Knoxville, Tennessee 37996, USA}
\affiliation{Institute of Advanced Materials and Manufacturing, The University of Tennessee, Knoxville, Tennessee 37996, USA\looseness=-1} 

\date{\today}

\begin{abstract}
\noindent{
The high-temperature superconducting cuprates host unidirectional spin- and charge-density-wave orders that can intertwine with superconductivity in non-trivial ways. While the charge components of these stripes have now been observed in nearly all cuprate families, their detailed evolution with doping varies across different materials and at high and low temperatures. We address this problem using non-perturbative determinant quantum Monte Carlo calculations for the three-band Hubbard model. Using an efficient implementation, we can resolve the model's fluctuating spin and charge modulations and map their evolution as a function of the charge transfer energy and doping. We find that the incommensurability of the charge modulations is decoupled from the spin modulations and decreases with hole doping, consistent with experimental measurements at high temperatures. These findings support the proposal that the high-temperature charge correlations are distinct from the intertwined stripe order observed at low-temperature and in the single-band Hubbard model. 
}
\end{abstract}

\maketitle

\vspace{1cm}

\noindent Understanding the anomalous normal-state of the high-temperature (high-$T_\mathrm{c}$) cuprates is one of the most fascinating and unresolved problems in condensed matter physics~\cite{Keimer2015Nature}. 
Strong electron-electron interactions govern this state, producing a highly correlated electron liquid that defies effective single-particle descriptions. In particular, underdoped cuprates exhibit intertwined order at low temperature $T$ \cite{FradkinRMP2015, Keimer2015Nature}, where combined unidirectional spin- and charge-density-waves (i.e., stripes) that can compete or cooperate with superconductivity in non-trivial ways. For example, stripe order can compete with superconductivity~\cite{TranquadaNature1995, Tranquada2020cuprate} or drive the formation of novel pair density waves, depending on the material~\cite{HamidianNature2016, DuNature2020, WangNC2021, ChenPNAS2022}. Charge-density-waves, associated with the charge component of the stripes~\cite{Choi2024universal}, have also observed across nearly all families of cuprates~\cite{LeePnas2022, Arpaia2021, CominARCMP2016}, sometimes extending up to as high as room temperature. Thus, understanding the origin of these novel spin and charge correlations is crucial to understanding the cuprate normal state and unraveling the mystery of cuprate superconductivity. 

The single-band Hubbard model is widely believed to be the low-energy effective model for the high-$T_\mathrm{c}$ cuprates~\cite{Zhang1988effective, Arovas2022hubbard}. The last decade has seen impressive advances in obtaining reliable numerical solutions to this model~\cite{HubbardRevComp2022}. For example, several numerical methods have obtained direct evidence for fluctuating \cite{HuangQM2018, MaiPNAS2022, Mai2023robust} or static \cite{ZhengScience2017, WietekPRX2021, XiaoPRX2023} stripe order in its strong coupling regime relevant to the cuprates. The calculations have also produced a general picture where the energy differences between competing metastable states are minimal, rendering the ground state highly sensitive to model parameters \cite{ZhengScience2017, JiangScience2019, ChungPRB2020}. At finite-$T$, \gls*{DQMC} and \gls*{METTS} have demonstrated an interlocking of spin and charge stripes in the single-band model~\cite{WietekPRX2021, HuangQM2018, MaiPNAS2022}, aligning with the Yamada relation~\cite{YamadaPRB1998} and consistent with experimental observations in doped La-based cuprates at low temperature~\cite{LeePnas2022, YamadaPRB1998, HuckerPRB2013, AbbamonteNphys2005, CroftPRB2014, MiaoNPJ2021,WenNC2019}. However, these simulations are incompatible with observations of a decreasing wave vector for the charge modulations with increasing hole-density in \gls*{YBCO}~\cite{AchkarPRL2012, Changnphys2012, GhiringhelliScience2012, BlackburnPRL2013, HuckerPRB2014, BlancoPRL2013, BlackburnPRL2013, BlancoPRB2014}, \gls*{Bi2201}~\cite{PengNatMat2018} and \gls*{Hg1201}~\cite{TabisPRB2017}. It has recently been proposed that this decrease may be a universal behavior of the high-temperature  charge fluctuations across all cuprates~\cite{LeePnas2022}. 

\begin{figure*}[ht]
    \centering
    \includegraphics[width=\textwidth]{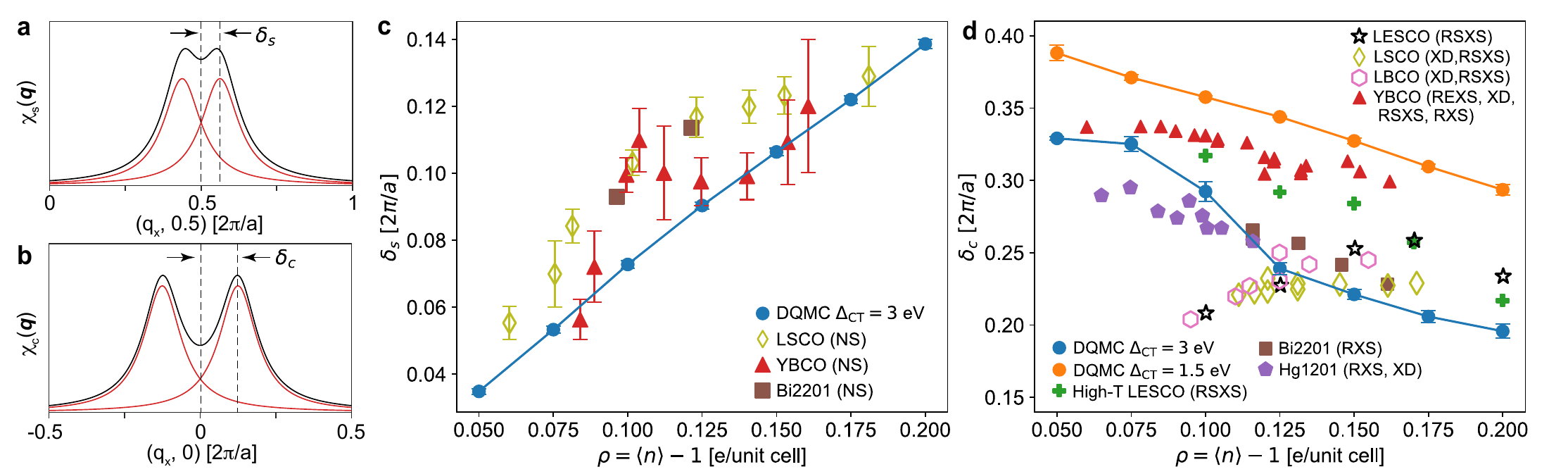}
    \caption{
    {\bf Spin and charge correlations in the three-band Hubbard model}. {\bf a} The incommensurate spin modulations 
    plotted in momentum space, where the spin susceptibility exhibits peaks at  
    ${\bf q} = (0.5\pm \delta_s,0.5)$ and 
    $(0.5,0.5\pm \delta_s)$ (measured in reciprocal lattice units). 
    {\bf b} The incommensurate charge modulations 
    plotted in momentum space, where the charge susceptibility exhibits peaks at 
    ${\bf q} = (\pm \delta_c,0)$ and 
    $(0,\pm \delta_c)$. 
    {\bf c} The doping dependence of $\delta_s$ determined from our \gls*{DQMC} simulations with charge transfer energy $\Delta_\mathrm{CT} = \epsilon_p-\epsilon_d = 3$ eV. Here, the results are compared with $\delta_s$ values measured in several cuprates including LSCO~\cite{YamadaPRB1998}, YBCO\cite{HeadingsPRB2011} and Bi2201\cite{EnokiPRL2013} using NS. {\bf d} The doping dependence of the charge 
    incommensurability determined from our \gls*{DQMC} simulations with different values of $\Delta_\mathrm{CT}$. The DQMC simulations are conducted at the inverse temperature $\beta=14~eV^{-1}$. The results are compared with the measured values extracted from X-ray scattering experiments on LESCO (RSXS~\cite{LeePnas2022}), LSCO (XD~\cite{CroftPRB2014, MiaoNPJ2021}, RSXS\cite{WenNC2019}), LBCO (XD~\cite{HuckerPRB2013}, RSXS~\cite{AbbamonteNphys2005}), YBCO (REXS~\cite{AchkarPRL2012}, XD~\cite{Changnphys2012, BlackburnPRL2013, HuckerPRB2014}, RSXS~\cite{GhiringhelliScience2012}, RXS~\cite{BlancoPRL2013, BlancoPRB2014}), Bi2201 (RXS~\cite{PengNatMat2018}) and Hg1201(RXS~\cite{TabisPRB2017}, XD~\cite{TabisPRB2017}). Abbreviation: NS, neutron scattering; RSXS, resonant soft X-ray scattering; REXS, resonant elastic X-ray scattering; XD, X-ray diffraction; RXS, resonant X-ray scattering.}
    \label{fig:Expcompare}
\end{figure*} 

Due to its coarse-grained nature, the single-band Hubbard model has limited ability to encapsulate specific-material characteristics among different cuprate families. It also cannot capture the crucial hybridization between the Cu and O orbitals, which govern excitations across the charge-transfer gap. These effects are instead described by the three-band Hubbard (or Emery) model, which explicitly includes the Cu $3d$ and O $2p$ degrees of freedom \cite{EmeryPRL1987}. 
Cluster dynamical mean-field theory and \gls*{DCA} studies of this model have observed a superconducting transition \cite{Weber2012scaling, WeberPRL2014, Mai2021orbital, Mai2021pairing} with a transition temperature $T_\mathrm{c}$ whose dependence on the charge transfer gap aligns with experimental data across various cuprates. Recent density matrix renormalization group studies on two-leg three-band ladders have also found evidence for a pair density wave superconducting ground state~\cite{Jiang2023pair, jiang2023pair_frustration}. Nonetheless, it remains an open question to what extent the spin and charge stripes in the three-band model are consistent with the experimental observations in the cuprates. So far, finite-$T$ \gls*{DQMC} calculations have observed fluctuating spin stripes \cite{HuangScience2017} but reported no corresponding charge component. Similarly, large-scale two-dimensional tensor network calculations have only detected a weak charge modulation that appears to vanish in the thermodynamic limit \cite{ponsioen2023superconducting}. Whether the three-band model supports a charge modulation in addition to the observed spin stripes remains an open question. 

Here, we present a detailed \gls*{DQMC} study of the fluctuating spin- and charge-correlations in the doped three-band Hubbard model. Advances in algorithms and computational power have allowed us to perform simulations at lower temperatures than the previous \gls*{DQMC} study~\cite{HuangScience2017}. We find evidence for fluctuating charge stripes in addition to the previously observed spin stripes, in both momentum and real space, for a range of hole densities and charge-transfer energies. Notably, we find that the incommensurability of the charge correlations  decreases with hole doping, opposite to the trend found for the single-band model but consistent with the high-$T$ behavior observed across different cuprate families~\cite{LeePnas2022}. In contrast, for the electron-doped case, we find that the density-dependent charge correlations in the three-band model are very similar to those observed in the single-band model. 
\\

\large
\noindent{\bf Results} 
\normalsize

\begin{figure*}[ht]
    \centering
    \includegraphics[width=0.9\textwidth]{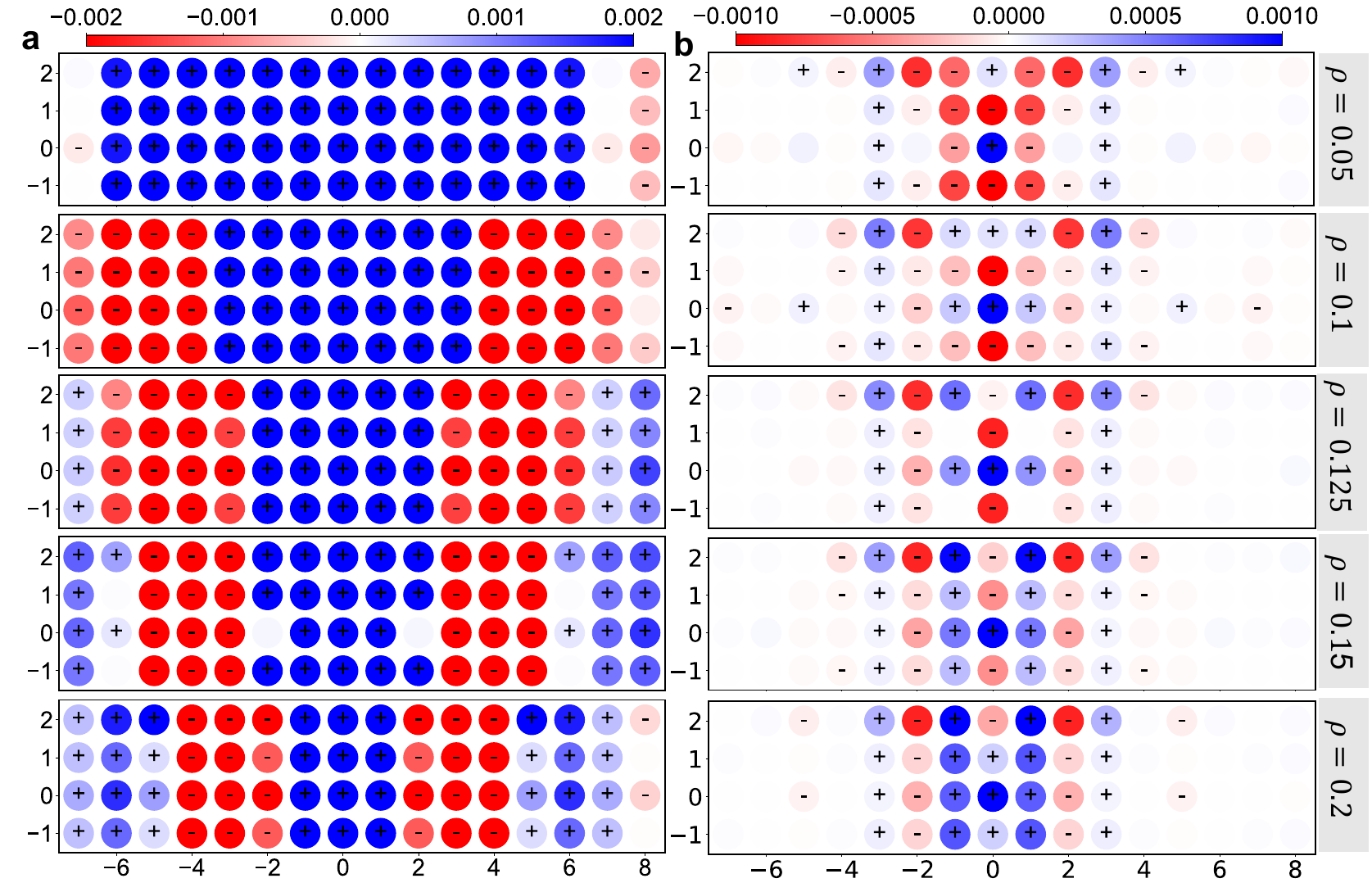}
    \caption{{\bf Static real-space Cu-Cu staggered spin and charge correlations at different hole densities}. The real-space Cu-Cu staggered spin ({\bf a}) and charge ({\bf b}) correlations are plotted at the lowest temperature ($\beta=14$) for varying hole densities as labeled. The signal on the sites labeled with $+$ or $-$ is above two $\sigma$, the Jackknife standard error.}
    \label{fig:RSsc}
\end{figure*}

\noindent Figure~\ref{fig:Expcompare} summarizes our main results. Our \gls*{DQMC} simulations reveal the presence of decoupled fluctuating spin and charge modulations in the three-band Hubbard model. The former are consistent with prior simulations~\cite{HuangScience2017}, while the latter are reported here for the first time. In real space, these correlations manifest as short-range correlations with a very short correlation length. In momentum space, the spin correlations appear as Lorentzian peaks centered around ${\bf q} = (0.5\pm \delta_s,0.5)$ in units of $2\pi/a$, as sketched in Fig.~\ref{fig:Expcompare}{\bf a}. Similarly, the charge correlations produce peaks in the density-density correlation function at ${\bf q} = (\pm \delta_c, 0)$, as illustrated in Fig.~\ref{fig:Expcompare}{\bf b}. If the two components form an intertwined stripe, then one expects $\delta_c = 2\delta_s$. [Note, our rectangular cluster geometry frustrates the equivalent peaks at $(0.5\pm \delta_s,0.5)$ and $(0,\pm \delta_c)$ that appear in simulations on square lattices~\cite{HuangScience2017}.]

Our simulations show that the incommensurability of the spin modulations $\delta_s$ increases with hole doping, as shown in Fig.~\ref{fig:Expcompare}{\bf c}. This behavior is consistent with prior results~\cite{HuangScience2017} and in quantitative agreement with experimental observations on a subset of cuprates. In contrast, the charge incommensurability $\delta_c$ decreases with hole doping as shown in Fig.~\ref{fig:Expcompare}{\bf d} for representative values of the bare charge transfer energy $\Delta_{\text{CT}}= 1.5$ and $3$ eV. The results for $\Delta_{\text{CT}}= 3$ eV are in excellent agreement with the wavevectors extracted from X-ray scattering measurements on \gls*{Bi2201}~\cite{PengNatMat2018} and \gls*{Hg1201}~\cite{TabisPRB2017} while the smaller value of $\Delta_{\text{CT}}$ agrees fairly well with the qualitative behavior observed in \gls*{YBCO}~\cite{AchkarPRL2012, Changnphys2012, GhiringhelliScience2012, BlackburnPRL2013, HuckerPRB2014, BlancoPRL2013, BlackburnPRL2013, BlancoPRB2014}. The opposite doping dependence of $\delta_s$ and $\delta_c$ indicate that the spin and charge modulations are not intertwined at this temperature. The behavior of the charge modulations presented here is consistent with the universal high temperature behaviors in experiment~\cite{LeePnas2022} and in contrast to the predictions of the single-band Hubbard model~\cite{MaiPNAS2022, HuangPRB2023}. We will now present the various measured quantities that lead to these results. Throughout, we work in hole language, where the number of excess holes in the unit cell is $\rho = \langle n \rangle -1$. 

\begin{figure*}[ht]
    \centering
    \includegraphics[width=\textwidth]{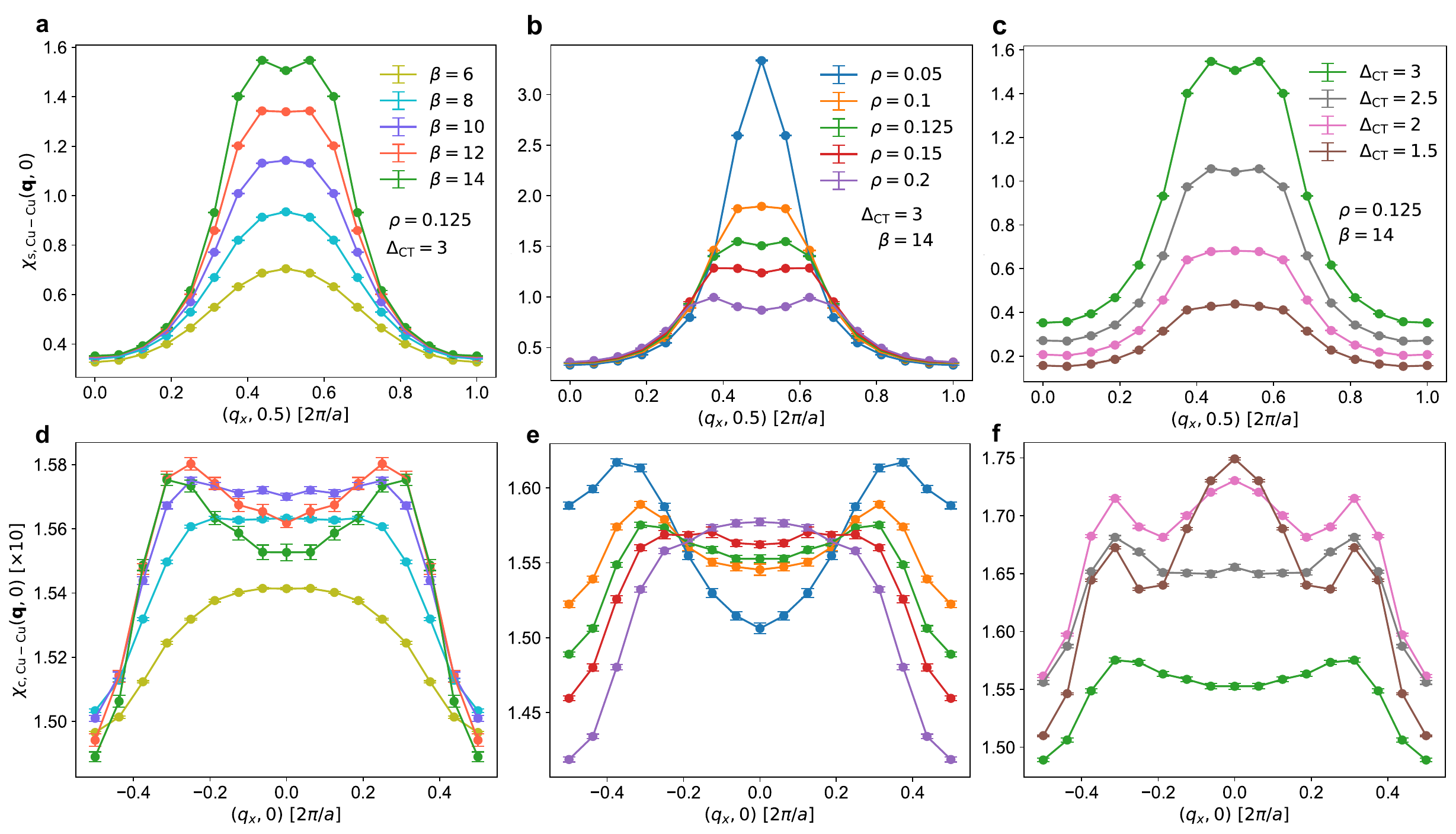}
    \caption{{\bf Static Cu-Cu spin and charge correlations in the three-band Hubbard model at varying temperatures, hole doping densities, and charge-transfer energies}. Panels {\bf a}, {\bf b} and {\bf c} show the static spin susceptibility $\chi_\text{s}({\bf q},0)$ for ${\bf q} = (q_x,0.5)$ at varying temperatures, hole densities, and charge-transfer energies, respectively, while fixing the other two parameters as labeled. Panels {\bf d}, {\bf e} and {\bf f} show the corresponding charge susceptibility $\chi_\text{c}({\bf q},0)$ along the ${\bf q} = (q_x,0)$  direction. The corresponding panels at the first and second rows share the same legend. These results are from DQMC simulations on a $16\times 4$ cluster.}
    \label{fig:vary}
\end{figure*}

Figure~\ref{fig:RSsc} plots the static Cu-Cu staggered spin-spin $\chi_\text{stag,s}({\bf r},\omega=0)$ and density-density $\chi_\text{c}({\bf r},\omega=0)$ correlation functions in real space. [Here, we include a ${\bf r}_i$-dependent staggered phase factor $\exp({\mathrm i}(\pi,\pi)\cdot {\bf r}_j)$ so that regions with the same blue or red color represent short-range antiferromagnetic domains but with a relative $\pi$-phase shift.]  Results are shown for several hole doping densities and obtained on $16\times4$ clusters with an onsite Hubbard interaction $U_d=6~\text{eV}$ on the Cu $3d$-orbitals, a fixed charge-transfer energy $\Delta_{\text{CT}} = 3$ eV, and inverse temperature $\beta = 14$ eV$^{-1}$. A clear unidirectional spin stripe is observed in Fig.~\ref{fig:RSsc}{\bf a} for $\rho > 0.1$ holes per unit cell, consistent with prior results~\cite{HuangScience2017}. The charge stripe pattern is also apparent in Fig.~\ref{fig:RSsc}{\bf b}; however, its overall amplitude and correlation length are much smaller than those for the spin correlations. This observation suggests that the charge modulations have a lower temperature scale compared with the spin modulations, consistent with predictions for the single-band model~\cite{MaiPNAS2022, HuangPRB2023} and the weak signals previously reported in the three-band model~\cite{ponsioen2023superconducting}. 

Due to their short correlation length at high temperatures, the fluctuating correlations are much easier to discern in momentum space, where they manifest as Lorentzian peaks rather than decaying exponentials. Fig.~\ref{fig:vary} summarizes the static Cu-Cu spin-spin $\chi_\text{s}({\bf q},\omega=0)$ and density-density $\chi_\text{c}({\bf q},\omega=0)$ correlation functions in momentum space. Here the spin susceptibility is plotted along the $(q_x,0.5)$ direction, while the charge susceptibility is plotted along the $(q_x,0)$ direction, as illustrated in Fig.~\ref{fig:Expcompare}{\bf a} and {\bf b}, respectively. 

Figures~\ref{fig:vary}{\bf a},{\bf d} plot the results as a function of temperature with a fixed hole-doping $\rho=1/8$ and $\Delta_{\text{CT}}=3~\text{eV}$. In this case, a double peak structure appears in the Cu-Cu spin correlations as the temperature is lowered, indicative of the formation of fluctuating spin stripe correlations. The formation of charge modulations is also apparent in Fig.~\ref{fig:vary}{\bf d} as the system cools down. While the amplitude of the peaks is smaller than their spin counterpart, the charge peaks can be discerned directly at lower temperatures without the need for curve fitting.

\begin{figure*}[ht]
    \centering
    \includegraphics[width=\textwidth]{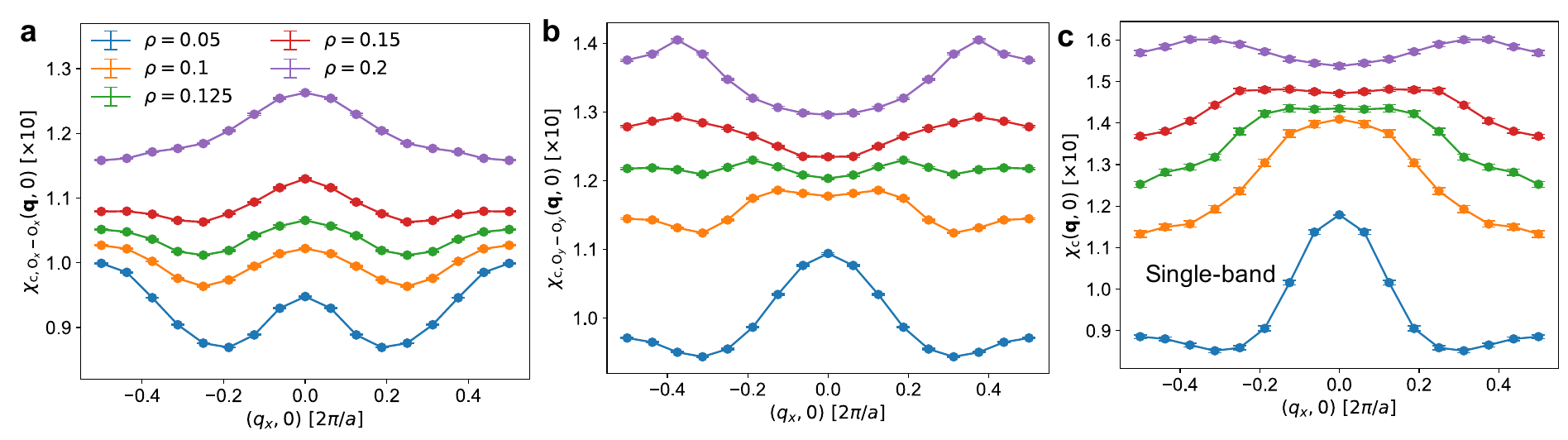}
    \caption{{\bf Static O-O correlations at different hole-doping densities in comparison with the single-band results}. Panel {\bf a} and {\bf b} show the static O$_x$-O$_x$ and O$_y$-O$_y$ charge susceptibility at $\beta=14$ eV$^{-1}$,$\Delta_{\text{CT}}=3$ eV, respectively, at different hole-doping densities $\rho$ for comparison with the DQMC charge correlation from the single-band Hubbard model ($t'/t=-0.25, U/t=6, \beta=4.5/t$) in panel {\bf c}. All panels share the same legend.}
    \label{fig:orbitals}
\end{figure*}

Figures~\ref{fig:vary}{\bf b} and {\bf e} examine the hole-density-dependence of the correlations at fixed $\Delta_\mathrm{CT} = 3$ eV and $\beta=14$ eV$^{-1}$. Here, the spin correlations evolve from a single antiferromagnetic peak centered at $(0.5,0.5)$ to incommensurate peaks at $(0.5\pm \delta_s,0.5)$ with increasing hole-doping, again consistent with previous studies~\cite{HuangScience2017} and the single-band Hubbard model~\cite{HuangQM2018, MaiPNAS2022, HuangPRB2023}. In the single-band case, however, the charge incommensurability $\delta_c$ is  increases commensurately with the spin correlations with hole doping~\cite{MaiPNAS2022, HuangPRB2023} (reproduced in Fig.~\ref{fig:orbitals}{\bf c}), consistent with an intertwined stripe. In contrast, we find that $\delta_c$ decreases with hole doping $\rho$ in the three-band model, as shown in Fig.~\ref{fig:vary}{\bf e} and summarized in Fig.~\ref{fig:Expcompare}{\bf c}.  Thus, there appears to be a fundamental difference between the spin- and charge-correlations of the two models at these temperatures, even though the single-band model correctly captures the pairing correlations of the three-band model~\cite{Mai2021orbital}. 

The charge transfer energy $\Delta_\text{CT}$ is a key parameter controlling the property variations across different families of cuprates~\cite{Ohta1991apex, AdlerRPP2019, WangScience2023}. Therefore, Figs.~\ref{fig:vary}{\bf c} and {\bf f} explore the evolution of the spin and charge correlations, respectively, as a function of $\Delta_{\text{CT}}$ at $1/8$-hole doping and $\beta = 14$ eV$^{-1}$. Decreasing $\Delta_{\text{CT}}$ enhances the charge susceptibility while suppressing the spin susceptibility, which also suggests the decoupling between charge and spin fluctuations at high temperatures. The double-peak stripe structure in the charge susceptibility persists as a 
function of $\Delta_\mathrm{CT}$, and is enhanced relative to a central ${\bf q} = 0$ peak as $\Delta_\mathrm{CT}$ decreases. (The latter also develops as $\Delta_{\text{CT}}$ decreases due to the transfer of holes from the O to Cu orbitals.) Interestingly, we also observe that the charge correlations have a non-monotonic dependence on the charge transfer energy, and appear to be strongest for $\Delta_\mathrm{CT} = 2$ eV. The overall dependence on $\Delta_{\text{CT}}$ indicates that the charge correlations observed here are robust over a wide range of charge transfer values, consistent with the ubiquitous presence of \gls*{CDW} correlations across different cuprates~\cite{CominReview}. The average value of the Fermion sign increases significantly as $\Delta_{\text{CT}}$ decreases, allowing us to reach lower temperatures. For example, the supplementary materials show results down to $\beta = 20$ eV$^{-1}$ for $\Delta_{\text{CT}} = 1.5$ eV, where the charge correlations become even more apparent. Therefore, we expect that these correlations will persist at lower temperatures.  

So far, we have focused on the Cu-Cu charge correlations, where the strongest signals are observed. The inter-orbital charge correlations (see supplement) are significantly weaker than the diagonal intra-orbital ones, presumably because we have neglected any Cu-O interactions. Figs.~\ref{fig:orbitals}{\bf a} and {\bf b} summarize the intra-orbital O$_\text{x}$-O$_\text{x}$ and O$_\text{y}$-O$_\text{y}$ charge correlations, respectively, along the $(q_x,0)$ direction as a function of hole-doping $\rho$ while fixing $\beta=14$ eV$^{-1}$ and $\Delta_{\text{CT}}=3$ eV. On a square lattice, one would expect the O$_\text{x}$-O$_\text{x}$ and O$_\text{y}$-O$_\text{y}$ correlations to differ when taken along the same momentum cut but to be identical for cuts rotated by $\pi/2$. Our $16\times4$ clusters introduce a slight anisotropy in the oxygen occupations (see supplement) and corresponding density-density correlations. Comparing to the Cu-Cu correlation in Fig.~\ref{fig:vary}{\bf e}, the O$_\text{x}$-O$_\text{x}$ and O$_\text{y}$-O$_\text{y}$ correlations shown in Fig.~\ref{fig:orbitals}{\bf a},{\bf b} bear a closer resemblance to the charge correlations obtained for single-band model as shown in Fig.~\ref{fig:orbitals}{\bf c}. In particular, the O$_\text{y}$-O$_\text{y}$ correlations are similar, with double peaks appearing and separating from each other as the doping increases. However, in real space, neither O$_\text{x}$-O$_\text{x}$ nor O$_\text{y}$-O$_\text{y}$ show a clear unidirectional stripe pattern (see supplement), indicating that the O correlations are still weak at this temperature. \\

\large
\noindent\textbf{Discussion} 
\normalsize 

\noindent The doping evolution of the Cu-Cu spin and charge modulations are qualitatively, and in some cases quantitatively, consistent with observations on a subset of the high-$T_\mathrm{c}$ cuprates. In particular, the evolution of the charge modulations with doping is in excellent agreement with many Cu $L$-edge resonant x-ray scattering probes, as summarized in Fig.~\ref{fig:Expcompare}{\bf d}, which are sensitive to the Cu charge density. 

Our results support the proposal that the high-$T$ charge correlations behave very differently from the low-$T$ intertwined stripe order that has been observed in several cuprate families~\cite{LeePnas2022}. While we can observe robust charge correlations down to $\beta = 20~\text{eV}^{-1}$ $(T = 580~\text{K})$ in some cases, it is unclear how the high-$T$ correlations connect to the low-$T$ correlations, if at all. At our simulation temperatures, there are a significant number of holes thermally excited across the charge transfer gap. The three-band Hubbard model can capture this charge transfer, while the single-band model cannot, leading to qualitatively different behavior. One possibility is that the freezing out of the charge transfer excitations will lock the charge modulations into the spin modulations, resulting in an intertwined order consistent with the single-band model. Constrained path quantum Monte Carlo calculations and other methods capable of bridging the high and low-temperature regimes may be be able to shed additional light on this problem.

Another outstanding issue is the relative strength of the spin and charge correlations. The single-band Hubbard model predicts~\cite{MaiPNAS2022, HuangPRB2023} that the spin modulations are significantly stronger than the charge modulations, while the situation is reversed in the actual materials~\cite{TranquadaNature1995, Tranquada2020cuprate, CominReview}. Our calculations show that the three-band model has a similar inverted hierarchy, consistent with the vanishing of charge correlations also inferred from 2D tensor network calculations \cite{ponsioen2023superconducting}. Accounting for this discrepancy may require the inclusion of additional (possibly density dependent) hopping pathways~\cite{Jiang2023density, yang2023recovery}, additional O $2p$ and Cu $3d$ – O $2p$ Coulomb interactions, or other couplings like the electron-phonon coupling~\cite{Karakuzu2022stripe, Chen2021anomalously}. \\

\large
\noindent\textbf{Methods}
\normalsize 

\noindent{\bf The Model} --- We studied the three-band Hubbard model for the CuO$_2$ plane. The Hamiltonian is 
\begin{align}\nonumber
    H&= \sum_{i,\sigma}(\epsilon_d - \mu)d^\dagger_{i,\sigma}d^{\phantom\dagger}_{i,\sigma}
    +\sum_{i,\alpha,\sigma} (\epsilon_p - \mu)p^\dagger_{i,\alpha,\sigma}p^{\phantom\dagger}_{i,\alpha,\sigma}\\\nonumber
    &+t_{pd} \sum_{i,j,\alpha} P_{i,j}^\alpha\left[d^\dagger_{i,\sigma}p^{\phantom\dagger}_{j,\alpha,\sigma}+\mathrm{H.c.}\right] \\\nonumber
    &+t_{pp} \sum_{j,j^\prime,\alpha,\alpha^\prime} Q_{j,j^\prime}^{\alpha,\alpha^\prime}\left[p^{\dagger}_{j,\alpha,\sigma}p^{\phantom\dagger}_{j^\prime,\alpha^\prime,\sigma}+\mathrm{H.c.}\right] \\
    &+U_d\sum_{i}d^\dagger_{i,\uparrow}d^{\phantom\dagger}_{i,\uparrow}d^\dagger_{i,\downarrow}d^{\phantom\dagger}_{i,\downarrow}
    \label{eq:H}
\end{align}
Here, $d^\dagger_{i,\sigma}$ ($d^{\phantom\dagger}_{i,\sigma}$) creates (annihilates) a spin $\sigma$ ($=\uparrow,\downarrow$) hole in the copper $d_{x^2-y^2}$ orbital at site $i$;  $p^\dagger_{i,\alpha,\sigma}$ ($p^{\phantom\dagger}_{i,\alpha,\sigma}$) creates (annihilates) a spin  $\sigma$ hole in the oxygen $p_\alpha$ ($\alpha = x,y$) orbital at site $j$; $t_{pd}$ and $t_{pp}$ are the magnitude of the nearest neighbour Cu-O and O-O hopping, respectively; $P_{i,j}^\alpha$ and $Q_{j,j^\prime}^{\alpha,\alpha^\prime}$ are the associated phase factors; $\epsilon_d$ and $\epsilon_p$ are the Cu and O site energies; $\mu$ is the chemical potential; and $U_{dd}$ is the on-site Cu Hubbard interaction. The orbital phase convention given by $P_{i,j}^\alpha$ and $Q_{j,j^\prime}^{\alpha,\alpha^\prime}$ is sketched in Fig.~1 of Ref.~\cite{Mai2021pairing}. We only consider the effect of on-site interaction in the Cu  $d_{x^2-y^2}$ orbital since including the on-site oxygen repulsion has a minor effect on the results but leads to a more severe sign problem~\cite{Mai2021orbital, Mai2021pairing, AcharyaPRX2018}.

Throughout, we fix $t_{pd} = 1.13$, $t_{pp} = 0.49$, $\epsilon_d = 0$, and $U_{dd} = 6$ (all in units of eV) and adjust the charge transfer $\Delta = \epsilon_p - \epsilon_d$ by changing the value of $\epsilon_p$. The value of $\mu$ is chosen to set the cluster's filling. 
\\

\noindent{\bf Determinant Quantum Monte Carlo} --- We solved Eq.~\eqref{eq:H} using unbiased \gls*{DQMC} simulations, as implemented in the \texttt{SmoQyDQMC.jl} package \cite{SmoQyDQMC}. We tuned the chemical potential $\mu$ for each parameter set to obtain the desired density within an accuracy of $O(10^{-4})$. We performed 8000 warm-up sweeps for each Markov chain, followed by 80000 measurement sweeps. Both equal- and unequal-time measurements were conducted once per sweep. We varied the number of Markov chains between 2500 and 12600 for each parameter set, depending on the severity of the sign problem and the intensity of the stripe pattern. The details of the average sign are provided in the supplemental information.\\

\noindent{\bf Measurements} --- To study the stripe behaviors, we measure the static ($\omega=0$) spin correlation function:
\begin{equation}
\chi_\mathrm{s,\alpha_1-\alpha_2}({\bf r},0) = \frac{1}{N}\sum_{\bf i}\int_0^\beta 
\langle \hat{S}^z_{\alpha_1,{\bf{r}+\bf{i}}}(\tau)\hat{S}^z_{\alpha_2,{\bf i}}(0)\rangle d\tau \label{cs1}
\end{equation}
and charge correlation function:
\begin{align}\nonumber
\chi_\mathrm{c,\alpha_1-\alpha_2}({\bf r},0) =&\frac{1}{N}\sum_{\bf i}\int_0^\beta [
\langle n_{\alpha_1,{\bf r}+{\bf i}}(\tau)n_{\alpha_2,{\bf i}}(0)\rangle 
\\&-\langle n_{\alpha_1,{\bf r}+{\bf i}}(\tau)\rangle \langle n_{\alpha_2,{\bf i}}(0)\rangle ] d\tau. \label{cs2}
\end{align}
where $\alpha_1,\alpha_2=\text{Cu}, \text{O}_\text{x}, \text{O}_\text{y}$ represent the orbital indices. We obtain the momentum-space spin and charge susceptibility shown in Fig.~\ref{fig:vary} and Fig.~\ref{fig:orbitals} through Fourier transforming Eq.~\eqref{cs1} and Eq.~\eqref{cs2}, respectively. For the real-space results in Fig.~\ref{fig:RSsc}, we plot the staggered spin correlation $\chi_\mathrm{s,\alpha_1-\alpha_2}({\bf r},0)(-1)^{r_x+r_y}$ and the charge correlation $\chi_\mathrm{c,\alpha_1-\alpha_2}({\bf r},0)$. \\

\large
\noindent{\bf Data Availability}
\normalsize 

\noindent The data supporting this study will be deposited in a public repository once the final version of the paper is accepted for publication. Until then, data will be made available upon reasonable requests made to S.J. (\href{mailto:sjohn145@utk.edu}{sjohn145@utk.edu}) or T.A.M. (\href{mailto:maierta@ornl.gov}{maierta@ornl.gov}).\\
 
\large
\noindent{\bf Code Availability}
\normalsize 

\noindent The \gls*{DQMC} code used for this project can be obtained at \url{https://github.com/SmoQySuite/SmoQyDQMC.jl}. \\

\large
\noindent{\bf References}
\normalsize 
\vspace{-1.6cm}
\bibliographystyle{naturemag}
\bibliography{references.bib}

\large
\noindent{\bf Acknowledgements}
\normalsize 

\noindent 
We thank T. P. Devereaux, E. Fradkin, and S. Kivelson for useful discussions. 
This work was primarily supported by the U.S. Department of Energy, Office of Science, Office of Basic Energy Sciences, under Award Number DE-SC0022311. This research used resources of the Oak Ridge Leadership Computing Facility, which is a DOE Office of Science User Facility supported under Contract No. DE-AC05-00OR22725. \\

\vspace{0.5cm}

\large
\noindent{\bf Author Contributions}
\normalsize 

\noindent P.M. performed the \gls*{DQMC} calculations for the three-band Hubbard model and analyzed the data. B. C.-S. developed the DQMC code. T.A.M. and S.J. supervised the project. All authors contributed to interpreting the data and writing the paper. \\

\large
\noindent{\bf Competing Interests}
\normalsize 

\noindent The authors declare no competing interests. \\

\end{document}


\clearpage

\onecolumngrid
\newpage

\renewcommand{\thefigure}{S\arabic{figure}}
\renewcommand{\theequation}{S\arabic{equation}}
\renewcommand{\thesection}{Supplementary Note \arabic{section}}

\setcounter{figure}{0}
\setcounter{equation}{0}
\setcounter{section}{0}

\title{Supplementary Information for ``Fluctuating charge-density-wave correlations in the three-band Hubbard model''}

\author{Peizhi Mai\orcidlink{0000-0001-7021-4547}}
\affiliation{Department of Physics and Anthony J Leggett Institute for Condensed Matter Theory, University of Illinois at Urbana-Champaign, Urbana, Illinois 61801, USA\looseness=-1}

\author{Benjamin Cohen-Stead\orcidlink{0000-0002-7915-6280}}
\affiliation{Department of Physics and Astronomy, The University of Tennessee, Knoxville, Tennessee 37996, USA}
\affiliation{Institute of Advanced Materials and Manufacturing, The University of Tennessee, Knoxville, Tennessee 37996, USA\looseness=-1} 

\author{Thomas A. Maier\orcidlink{0000-0002-1424-9996}}
\affiliation{Computational Sciences and Engineering Division, Oak Ridge National Laboratory, Oak Ridge, Tennessee, 37831-6494, USA\looseness=-1}

\author{Steven Johnston\orcidlink{0000-0002-2343-0113}}
\affiliation{Department of Physics and Astronomy, The University of Tennessee, Knoxville, Tennessee 37996, USA}
\affiliation{Institute of Advanced Materials and Manufacturing, The University of Tennessee, Knoxville, Tennessee 37996, USA\looseness=-1} 

\date{\today}

\maketitle

\onecolumngrid

\section{The average value of the Fermion sign and the orbital occupations}

\begin{figure*}[hb!]
    \centering
    \includegraphics[width=0.5\textwidth]{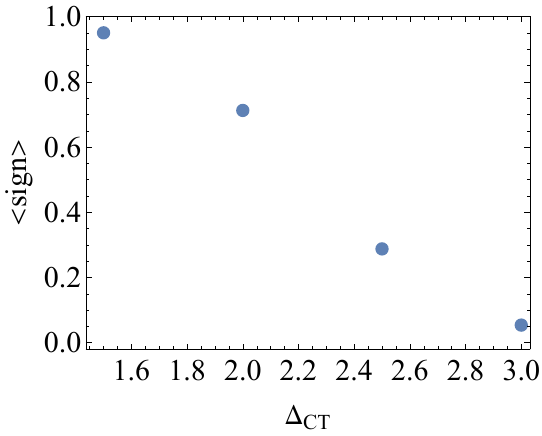}
    \caption{The average sign as a function of charge transfer energy at hole doping density $\rho=0.125~(\langle n\rangle =1.125$) and $\beta=14$ eV$^{-1}$.}
    \label{supfig:signvsdelta}
\end{figure*}

Figure.~\ref{supfig:signvsdelta} plots the average value of the Fermion sign in our simulations as a function of charge transfer energy and fixed $\beta = 14$ eV$^{-1}$ and $\rho=0.125$ hole doping $(\langle n\rangle =1.125$). We find that the average value of the sign increases significantly as $\Delta_\mathrm{CT}$ decreases. We are thus able to explore much lower temperatures for smaller values of $\Delta_{\text{CT}}$, as shown in Figs.~\ref{supfig:d1p5} {\bf a} and {\bf c}.

\begin{figure*}[ht]
    \centering
    \includegraphics[width=\textwidth]{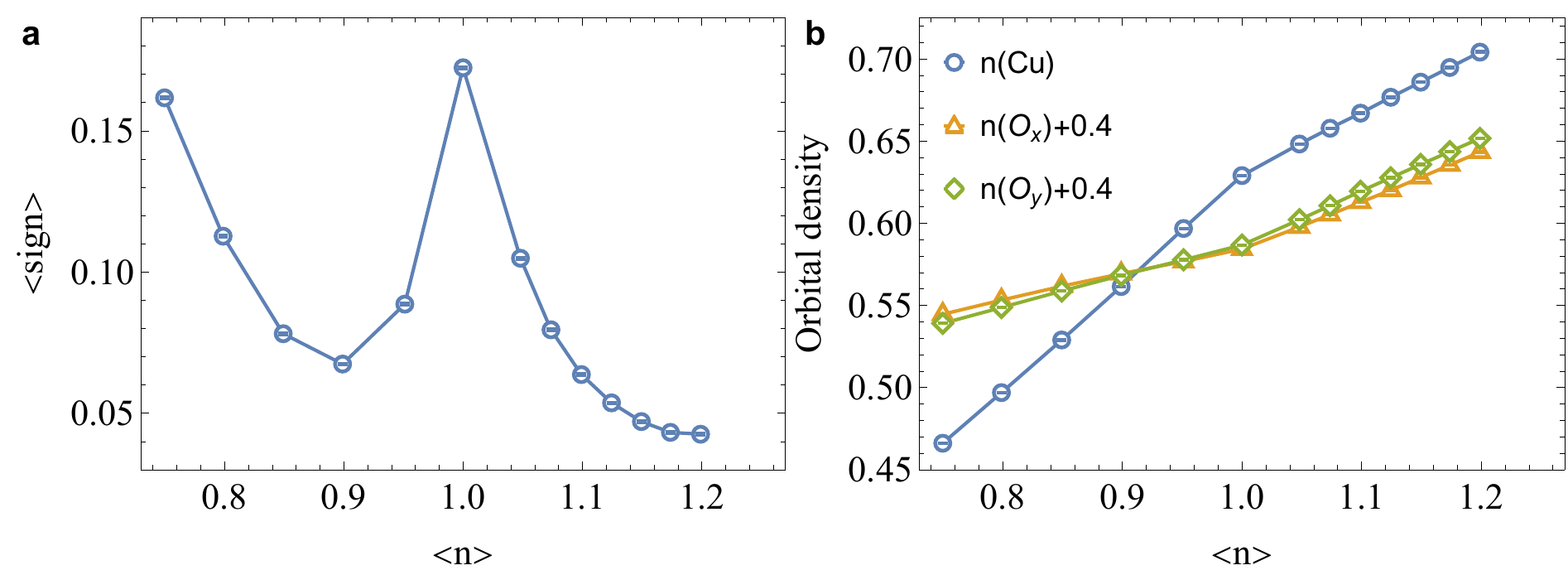}
    \caption{{\bf (a) The averaged Fermion sign and (b) orbital density as a function of total hole density at $\Delta_{\text{CT}}=3$ eV and the lowest temperature $\beta=14$ eV$^{-1}$}.}
    \label{supfig:denNsign}
\end{figure*}

Figure~\ref{supfig:denNsign}{\bf a} plots the average Fermion sign as a function of hole density $\langle n\rangle$ for fixed $\Delta_{\text{CT}}=3$ eV and $\beta=14$ eV$^{-1}$. The results are particle-hole asymmetric, and the average value of the sign improves in the vicinity of half-filling, consistent with previous reports~\cite{KungPRB2016}. 

Figure~\ref{supfig:denNsign}{\bf b} plots the average orbital occupations as a function of the total hole density. At this interaction strength ($U_{d}=6$ eV) and charge transfer energy ($\Delta_{\text{CT}}=3$ eV), most of the doped holes go to the O orbitals; however, the Cu orbitals still host a considerable proportion of the doped holes due to the strong hybridization between the Cu and O orbitals. 

\section{Additional results for the spin and charge correlations}
\begin{figure*}[ht!]
    \centering
    \includegraphics[width=\textwidth]{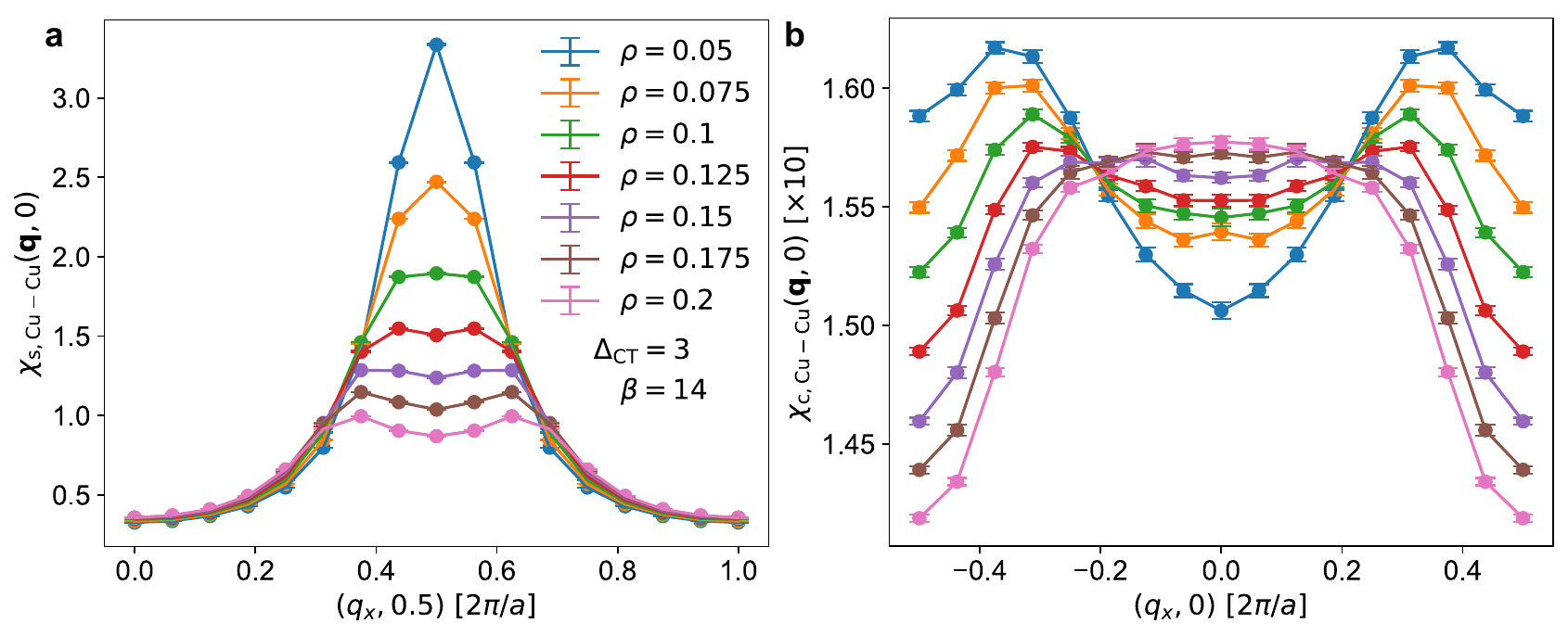}
    \caption{{\bf The doping dependence of the spin and charge correlations.}
    The static Cu-Cu ({\bf a}) spin and ({\bf b}) charge correlations in the three-band Hubbard model at varying hole doping densities with $\Delta_{\text{CT}}=3$ eV and $\beta=14$ eV$^{-1}$. The remaining model parameters are identical to those used in the main text.}
    \label{supfig:allden}
\end{figure*}

Figure~\ref{supfig:allden} plots the static Cu-Cu spin and charge susceptibilities for a hole density $\rho$ ranging from $0.05$ to $0.2$ in increments of $0.025$. Here, we have fixed $\Delta_{\text{CT}}=3$ eV and $\beta=14$~eV$^{-1}$. As $\rho$ increases, the \gls*{AFM} peak in the spin susceptibility splits into two peaks that continue to separate from one another as the doping increases. This behavior indicates that the wave vector of the spin correlations increases with hole-doping, consistent with the results in the main text and prior studies~\cite{HuangScience2017}. Conversely, the amplitude of the peaks in the charge susceptibility decreases with hole doping, and their locations shift continuously toward the zone center as more holes are introduced. 

\begin{figure*}[bt!]
    \centering
    \includegraphics[width=0.9\textwidth]{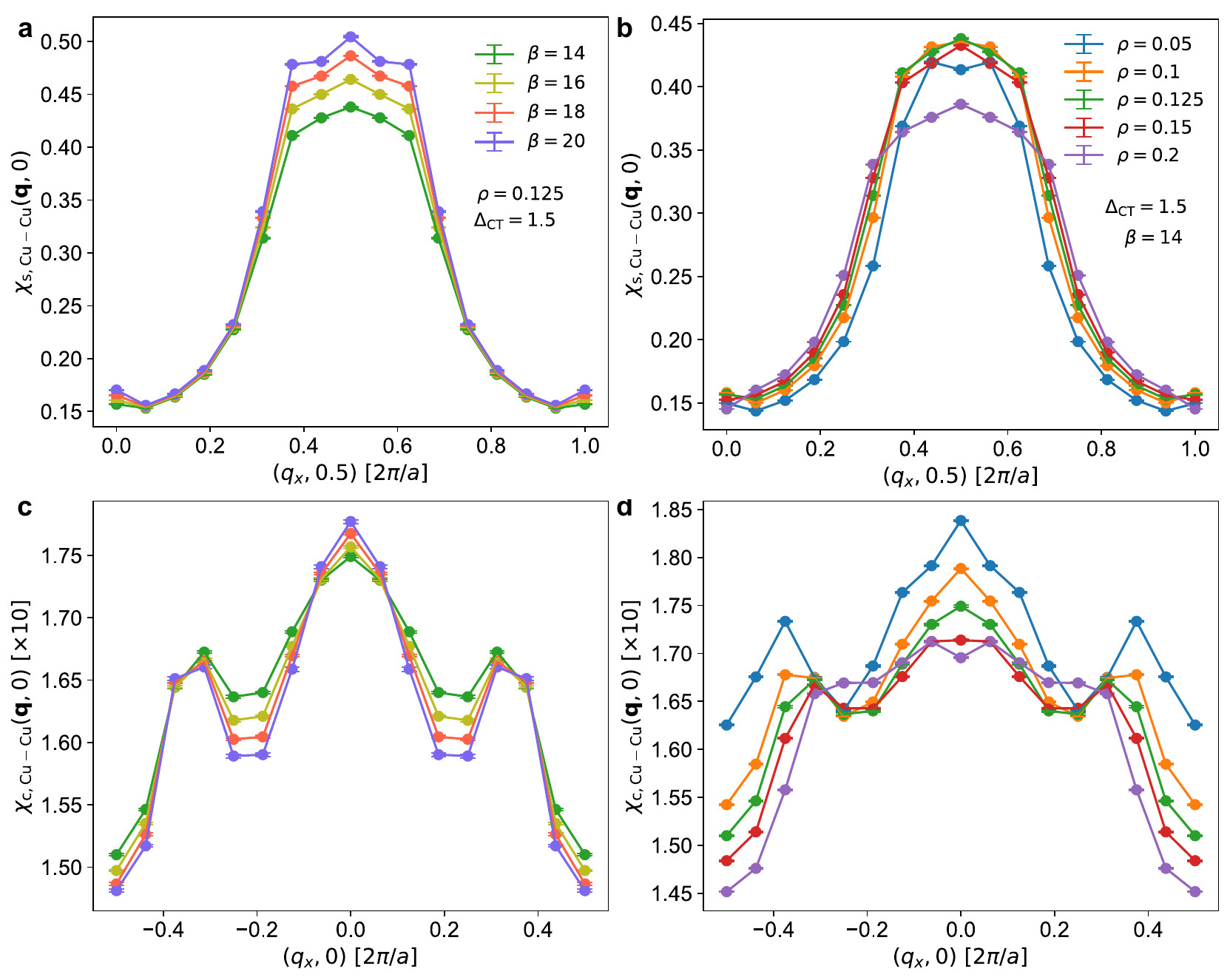}
    \caption{{\bf Static Cu-Cu spin and charge correlations in the three-band Hubbard model at varying temperatures and hole doping densities with $\Delta_{\text{CT}}=1.5$}. Panels {\bf a} and {\bf b} show the static spin $\chi_\text{s}({\bf q},0)$ susceptibility ${\bf q} = (q_x,0.5)$ (in units of $2\pi/a$) direction at varying temperatures and hole densities respectively, while fixing the other two parameters as labeled. Panels {\bf c} and {\bf d} show the corresponding charge susceptibility $\chi_\text{c}({\bf q},0)$ along the ${\bf q} = (q_x,0)$  direction. The corresponding panels at the first and second rows share the same legend. These results are from DQMC simulations on a $16\times 4$ cluster.}
    \label{supfig:d1p5}
\end{figure*}

The alleviation of the sign problem at smaller charge transfer energies $\Delta_{\text{CT}}$ (shown in \figdisp{supfig:signvsdelta}) makes it possible to explore lower temperatures in some cases. To this end, Figs.~\ref{supfig:d1p5}{\bf a} and {\bf c} show the Cu-Cu spin and charge susceptibilities, respectively, at a function of temperatures $T<1/14$ eV with $\Delta_{\text{CT}}=1.5$~eV and $\rho=0.125$. In this case, the amplitude of the peaks in the spin susceptibility increases as the temperature decreases while their incommensurability $\delta_s$ does not appear to change. On the other hand, the charge susceptibility has a three-peak structure, with a dominant central peak accompanied by two sub-dominant incommensurate peaks, which all become sharper as the system cools down. 

Figs.~\ref{supfig:d1p5}{\bf b} and {\bf d} show the hole doping density dependence of the spin and charge susceptibility, respectively, fixing $\beta=14$ eV$^{-1}$ and $\Delta_{\text{CT}}=1.5$ eV. For these parameters,  incommensurability in the spin susceptibility $\delta_s$ remains relatively small at all doping levels. Because of this, the two peaks merge into a broadened single peak. 
The incommensurate peaks in the charge susceptibility also exist for all hole doping levels and show a similar doping dependence as the results for $\Delta_{\text{CT}}=3$ eV demonstrated in the main text.

\begin{figure*}[ht!]
    \centering
    \includegraphics[width=0.6\textwidth]{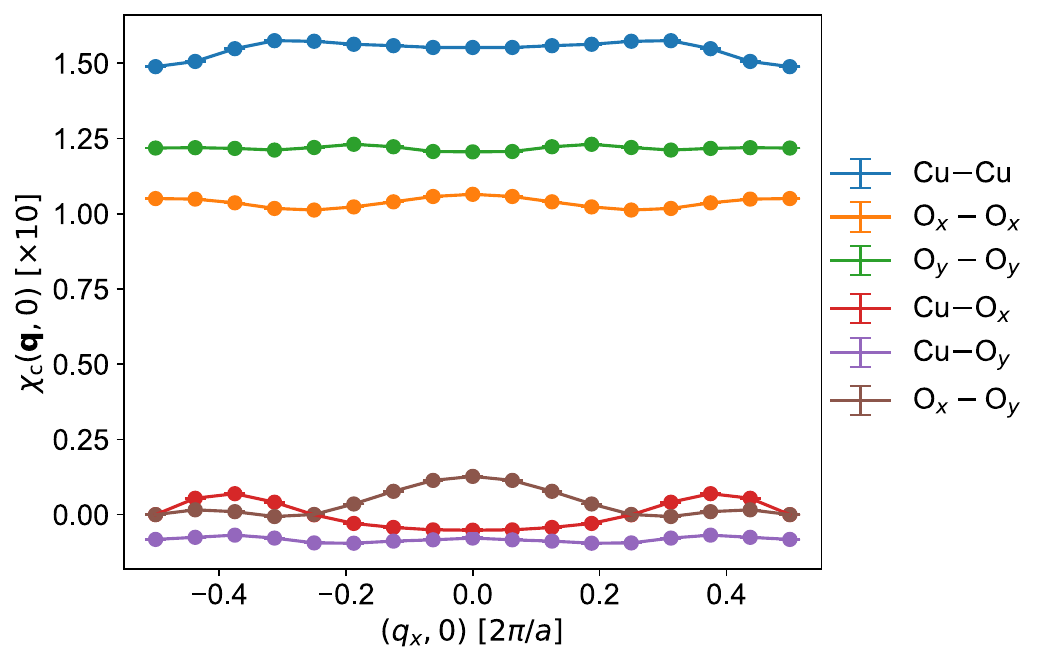}
    \caption{{\bf Different orbital components of static charge susceptibility at hole doping density $\rho=0.125~(\langle n\rangle =1.125$) and $\beta=14$ eV$^{-1}$}.}
    \label{supfig:interorbit}
\end{figure*}

Finally, Fig.~\ref{supfig:interorbit} shows all orbital components of the charge susceptibility at $\rho=0.125$, $\beta=14$ eV$^{-1}$, and $\Delta_{\text{CT}}=3$ eV. The inter-orbital correlations are negligible compared to the intra-orbital ones, probably due to the lack of inter-orbital Hubbard interactions. For this reason, the main text focuses on the intra-orbital components, particularly the Cu-Cu component, which has the most prominent signal.

\section{Oxygen-oxygen charge correlations in real space}

\begin{figure*}[ht!]
    \centering
    \includegraphics[width=\textwidth]{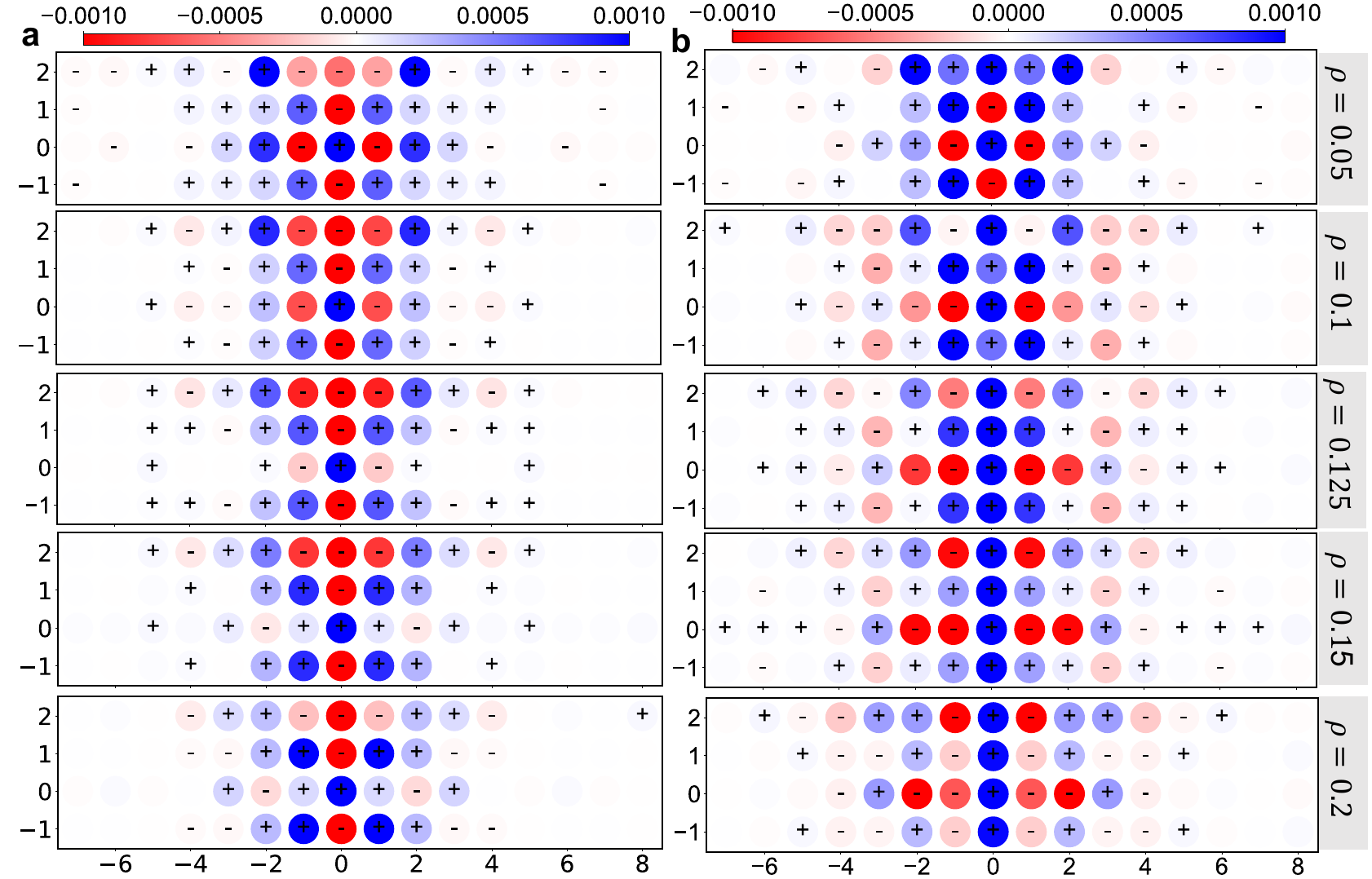}
    \caption{{\bf Static real-space O-O charge correlations at different hole densities}. The real-space ({\bf a}) O$_x$-O$_x$ and ({\bf b}) O$_y$-O$_y$ charge correlations are plotted at the lowest temperature ($\beta=14$ eV$^{-1}$) and $\Delta_{\text{CT}}=3$ eV, varying hole densities as labeled. The plotted signal labeled with $+$ or $-$ are above two $\sigma$, the Jackknife standard error.}
    \label{supfig:rsOO}
\end{figure*}

Figure~\ref{supfig:rsOO} shows the real-space O$_x$-O$_x$ (panel {\bf a}) and O$_y$-O$_y$ (panel {\bf b}) components of the charge susceptibility at different hole doping levels with $\beta=14$ eV$^{-1}$ and $\Delta_{\text{CT}}=3$~eV. Points labeled with a $\pm$ sign indicate data points where the Monte Carlo mean is $>2\sigma$ from zero, where $\sigma$ is the Jackknife standard error. 
While there are points where the signal is greater than the error bars, the overall correlations are weak and do not present any clear uni-directional stripe-like pattern, especially compared to the Cu-Cu component shown in Fig.~2 of the main text. 

\section{Additional results for electron doping}
\begin{figure*}[ht]
    \centering
    \includegraphics[width=\textwidth]{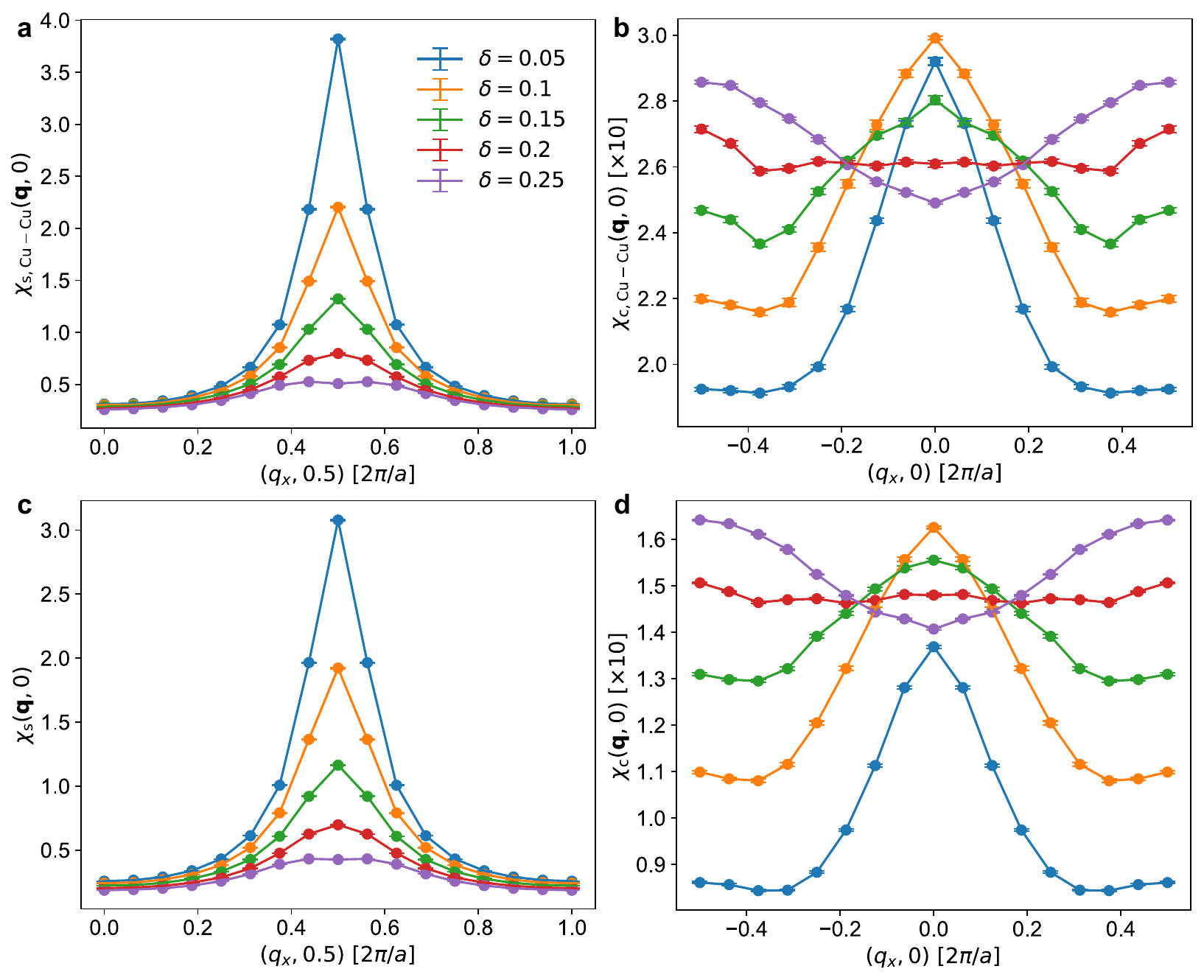}
    \caption{{\bf Static spin and charge correlations at varying doped electron densities in the three- and single-band Hubbard model}. The static Cu-Cu spin ({\bf a}) and charge ({\bf b}) correlations from the three-band Hubbard model are obtained at $\Delta_{\text{CT}}=3$ eV and $\beta=14$ eV$^{-1}$, while the spin ({\bf c}) and charge ({\bf d}) correlations from the single-band Hubbard model are obtained at $t'/t=0.25$, $\beta=4.5$. Both simulations are conducted in a 16$\times$4 cluster (unit cell) in each model. All panels share the same legend. }
    \label{supfig:edoped}
\end{figure*}

Figures~\ref{supfig:edoped}{\bf a} and {\bf b} show the static spin and charge susceptibilities on the electron-doped side of the phase diagram. Here, we fix $\Delta_{\text{CT}}=3$ eV and $\beta=14$ eV$^{-1}$. Unlike the hole-doped case, the electron-doped density-dependent line shapes for both spin and charge susceptibility are very similar to the single-band results plotted in \figdisp{supfig:edoped}{\bf c} and {\bf d}. This observation supports the notion that the single-band Hubbard model captures the low-energy spin and charge correlations of the three-band Hubbard model and the electron-doped cuprates~\cite{Mai2023robust}. 

\bibliography{references}

%% file: acronyms.tex
\newacronym{DQMC}{DQMC}{determinant quantum Monte Carlo}
\newacronym{AFM}{AFM}{antiferromagnetic}
\newacronym{QMC}{QMC}{quantum Monte Carlo}
\newacronym{DCA}{DCA}{dynamical cluster approximation}
\newacronym{DMRG}{DMRG}{density matrix renormalization group}
\newacronym{DMFT}{DMFT}{dynamical mean-field theory}
\newacronym{CDW}{CDW}{charge-density-wave}
\newacronym{SDW}{SDW}{spin-density-wave}
\newacronym{RIXS}{RIXS}{resonant inelastic x-ray scattering}
\newacronym{STM}{STM}{scanning tunneling microscopy}
\newacronym{YBCO}{YBCO}{YBa$_2$Cu$_3$O$_{6+\delta}$}
\newacronym{Bi2212}{Bi2212}{Bi$_2$Sr$_2$CaCu$_2$O$_{8+\delta}$ }
\newacronym{Bi2201}{Bi2201}{(Bi,Pb)$_{2.12}$Sr$_{1.88}$CuO$_{6+\delta}$ }
\newacronym{Hg1201}{Hg1201}{HgBa$_2$CuO$_{4+\delta}$}
\newacronym{METTS}{METTS}{minimally-entangled typical thermal states}
\newacronym{LESCO}{LESCO}{La$_{1.8−x}$Eu$_{0.2}$Sr$_x$CuO$_4$}
\newacronym{LSCO}{LSCO}{La$_{1-x}$Sr$_x$CuO$_4$}
\newacronym{LBCO}{LBCO}{La$_{1-x}$Ba$_x$CuO$_4$}